\documentclass[twocolumn,showpacs,amsmath]{revtex4}

\usepackage[dvips]{graphicx}
\usepackage{epsfig,euscript}

\begin{document}
\title{BaVS$_3$ probed by V L edge X-ray absorption spectroscopy}
\author{V. Ilakovac$^{1,2,*}$, 
N. B. Brookes$^{3}$, J. Criginski Cezar$^{3}$,  P. Thakur$^{3}$, V. Bisogni$^{3}$, 
C. Dallera$^{4}$, \\
G. Ghiringhelli$^{4}$, L. Braicovich$^{4}$, 
S. Bernu$^{5}$, 
H. Berger$^{6}$, L. Forr\'o$^{6}$, A. Akrap$^{6}$, C. F. Hague$^{1}$}
\email{vita.ilakovac-casses@upmc.fr}

\affiliation{$^1$Universit\'e Pierre et Marie Curie, CNRS UMR 7614, LCP-MR, 
F-75231 Paris, France\\
$^2$Universit\'e de Cergy-Pontoise, F-95031 Cergy-Pontoise, France\\
$^3$ESRF, B.P. 220, 38043 Grenoble Cedex, France\\
$^4$Politecnico di Milano, Piazza Leonardo da Vinci 32, 20133 Milano, Italy\\
$^5$Laboratoire de Physique de Solides, CNRS-UMR 8502, Universit\'e Paris-Sud, 
Batiment 510, 91405 Orsay Cedex, France\\
$^6$Ecole Polytechnique Federale de Lausanne CH-1015, Switzerland}

\date{\today}

\begin{abstract}
Polarization dependent vanadium L edge X-ray absorption spectra of BaVS$_3$ 
single crystals are measured in the four phases of the compound.
The difference between signals with the polarization \textbf{E}$\perp$\textbf{c} 
and \textbf{E}$\parallel$\textbf{c} (linear dichroism) changes with temperature. 
Besides increasing intensity of one of the maxima, 
a new structure appears in the pre-edge region below the metal-insulator transition. 
More careful examination brings to light that the changes start already  with 
pretransitional charge density wave fluctuations. 
Simple symmetry analysis suggests that the effect is related to rearrangements in $E_{g}$ 
and $A_{1g}$ states, and is compatible with the formation of 
four inequivalent V sites along the V-S chain. 
\end{abstract}


\maketitle

\section{\label{sec:Intro}Introduction}
Complex materials often display rich phase diagrams where different types of 
instabilities lead to unusual properties related to various 
ordering phenomena. Understanding the interplay between lattice, charge, spin and orbital 
degrees of freedom is one of most challenging aspects of 
correlated electron systems.

Barium vanadium sulfide, BaVS$_3$, is a particularly intriguing 
material because its structural, electrical and magnetic properties
point to a combination of quasi-1D and 3D behavior. 
Its room temperature structure consists of hexagonal packing 
of quasi 1D-chains of face-sharing VS$_6$ octahedra directed along the 
\textbf{c} axis \cite{Gardner}.
The ratio of inter-chain to intra-chain V-V distances being 
greater than 2, a high conductivity anisotropy ratio 
($\sigma_c$/$\sigma_a$) can be expected.
Surprisingly, $\sigma_c$/$\sigma_a$  is only about three \cite{Mihaly}. 
Lowering the temperature induces a zig-zag deformation of the V chains 
at $T_S$ = 240~K and reduces the crystal symmetry from hexagonal to 
orthorhombic \cite{Gardner}, but does not significantly change its 
conductive and magnetic properties. In both phases, it is a paramagnetic metal.
Below $T_{MI}$ = 69~K it becomes a paramagnetic insulator (PI). 
The metal-insulator (MI) transition is driven by a Peierls instability \cite{Fagot1}, 
accompanied by a change to a monoclinic structure \cite{Inami}.
At $T_N$ = 31 K, it undergoes a third phase transition to
the lowest temperature (ground) state which has an incommensurate 
antiferromagnetic (AF) order in the (\textbf{a},\textbf{b}) plane \cite{H_NakamuraJPSJ}. 
Recent resonant magnetic x-ray scattering reveal additional incommensurability 
along the \textbf{c} axis in the AFI ground state, while the 
supporting time-dependent density functional theory simulations indicate that 
the spins lie within the (\textbf{a},\textbf{b}) plane and are polarized 
along the monoclinic \textbf{a} axis \cite{Leininger}. 

Diffuse X-ray scattering experiments \cite{Fagot1} 
have demonstrated that the MI transition is driven by a 1D instability of
the electron gas. It leads to the formation of charge density waves (CDW) at the
critical wave vector \boldmath$q$$_{c}$ = 2\boldmath$k$$_F$, preceded by pretransitional fluctuations.
But the physical properties of BaVS$_3$ are not those of 
a standard quasi-1D system. 
First, the wave vector $q_{c}$\unboldmath = 0.5~\textbf{c*} (\textbf{c*} stands for 
the reciprocal lattice vector related to \textbf{c}) indicates that only one of 
the two $d$ electrons per unit cell ($uc$) participates directly in the CDW.
On the other hand, instead of a typical Pauli behavior of a 1D 
electron gas in the metallic phase, it has a Curie-Weiss behavior 
with an effective magnetic moment of $\mu$ = 1.2 $\mu_{B}$ corresponding to
approximately one localized spin per two vanadium sites \cite{Mihaly}. 
All these properties point to an unusual situation with $one$ electron per $uc$ 
delocalized along the \textbf{c*} direction driving the CDW formation. 
The $other$ is localized and responsible for the magnetic behavior.

Local density approximation (LDA) calculations 
indicate an interplay between two different types of electron states at the 
Fermi level: two narrow $E_g$ bands, and one dispersive band 
with mainly $A_{1g}$ character extending along the \textbf{c*} 
direction \cite{Mattheiss,Whangbo}. 
But the LDA filling of the $A_{1g}$ band is almost complete 
($n_{A_{1g}}$ $\approx$ 1.90 per $uc$), 
which overestimates the itinerant character, and is incompatible 
with the above-mentioned experimental findings. Dynamical mean-field theory
(DMFT) calculations show that strong enough charge correlation and exchange 
effects can bring $n_{A_{1g}}$ down to $one$ per $uc$ \cite{Lechermann}. 
\begin{table}
\begin{center}
\begin{tabular}{lllll}
\hline
phase & AFI & mPI & oPM & hPM \\ 
$T_c$ & $T_N$= 31~K & $T_{MI}$= 69~K & $T_S$= 240~K &  \\ 
structure & monoclinic & monoclinic & orthorhombic & hexagonal \\ 
space group & $Im$ & $Im$ & $Cmc2_1$ & $P6_3mmc$ \\ 
point group & $m$ (C$_{1h}$) & $m$ (C$_{1h}$) & $mm2$ (C$_{2v}$) & 
$\overline{3}m$~(D$_{3d}$)\\
N$_V$ (N$_{ineq.}$) & 4 (4) & 4 (4) & 2 (1) & 2 (1) \\
magnetism & AF & para & para & para \\
transport & insulator & insulator & metal & metal \\ 
\hline
\end{tabular}
\end{center}
\caption{\label{tab_phases} 
Physical properties of BaVS$_3$ in its four phases: 
Anti-Ferromagnetic Insulator (AFI), monoclinic Paramagnetic Insulator (mPI), 
orthorhombic (oPM), and hexagonal (hPM) Paramagnetic Metal. 
$T_c$ stands for critical temperature, and N$_V$ (N$_{ineq.}$) 
the number of (inequivalent) V atoms per unit cell.}
\end{table}

Both types of electron states, delocalized (charge degrees of freedom), 
and localized (magnetic properties), 
are modified at the MI transition.
A charge gap of about 50~meV \cite{Mihaly} and a 
spin gap of 10-20~meV \cite{NakamuraPRL} are opened at $T_{MI}$, 
even if the spin degrees of freedom quench only in the AFI phase. 
But the mechanism of their interplay is not yet completely understood. 

We have explored four BaVS$_3$ phases in single crystals by polarization 
dependent x-ray absorption spectroscopy (XAS) at the V L edges. 
After describing the experimental means for obtaining very high resolution 
contamination-free data we present the results and then discuss how the 
polarization dependence of the XAS data provides new insight into the subtle 
electronic structure changes that take place across a temperature range that 
covers all the phases of BaVS$_3$. 

\section{\label{sec:Exp}Experiment}
Single crystals of BaVS$_3$ were grown by the tellurium flux method \cite{Berger}. 
Two samples from different batches were cleaved $in$-$situ$ along a plane 
parallel to the \textbf{c}-axis. After the cleaving at a pressure of 10$^{-9}$~mbar, 
they were transferred via a fast insertion lock to the analysis chamber at the base 
pressure of 4 x 10$^{-10}$~mbar. The experiments were performed using the helical 
undulator Dragon beam line ID08 at the European Synchrotron Radiation Facility 
(ESRF). At the V L$_3$ edge ($\approx$ 514~eV) the resolution was $\approx$ 100~meV. 
The incident light was normal to the sample surface and linearly polarized 
either parallel to the sample \textbf{c}-axis, or perpendicular to it. 
We periodically checked for energy shifts coming from the monochromator. 
The stability of the photon energy is estimated to be better than 40~meV.
The V L edge XAS signal was measured in the total electron yield mode 
using the drain current from the sample holder.

\begin{figure}
\begin{center}
\includegraphics [width=8.5cm,angle=0]{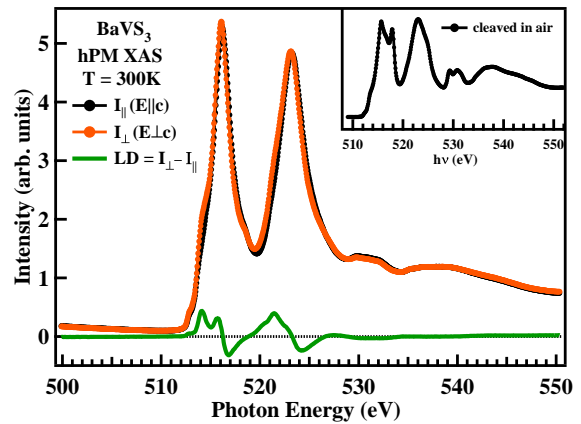}
\caption{\label{fig:XAS_pol_dep} (Color online) 
Paramagnetic Metallic phase (T = 300~K) 
XAS spectra of the $in$-$situ$ cleaved sample 
for the two polarizations of the incoming light:
\textbf{E}$\perp$\textbf{c} (red dots) and \textbf{E}$\parallel$\textbf{c} (black dots).
The linear dichroism 
(LD  = I$_{\perp}$-I$_{||}$) is also shown in the same graph (green line). 
The inset shows the spectra of a sample cleaved in air where the oxygen 
contamination is clearly visible above 530 eV. Extra features are observed at the V L$_3$ edge 
and  the L$_2$ peak is broadened.}
\end{center}
\end{figure}

\section{\label{sec:Res}Results}
V L$_{2,3}$ XAS results for BaVS$_3$ at room temperature for the 
two polarizations of the incoming light, 
\textbf{E}$\perp$\textbf{c} and 
\textbf{E}$\parallel$\textbf{c} are presented in Fig.~\ref{fig:XAS_pol_dep}. 
The spectra are normalized so as to obtain the same difference in amplitude below 
the L$_3$ edge and at 534~eV. 
They are separated by the 2$p$ spin orbit coupling, 
into L$_3$ ($h\nu$ = 512-519~eV) and 
L$_2$ ($h\nu$ = 519-528~eV) regions. 
The V L$_3$ and L$_2$ maxima are at 516.2~eV and 523.2~eV, respectively, 
i.e. 7~eV apart.
Their overall shape is closer to the spectra of vanadium 
metal \cite{Ilakovac} than to the spectra of vanadium oxides  
that tend to be more structured 
(see for example VO$_2$ \cite{Abbate} or V$_2$O$_3$ \cite{Park2000}). 

The O K XAS signal lies in the 530-545~eV region. 
By comparing with the inset in Fig.~\ref{fig:XAS_pol_dep} 
it can be seen that the surface contamination is negligibly small. 
In particular the hump at 540~eV is very weak while the feature at 531 eV 
may even be intrinsic to the BaVS$_3$ spectrum. 
A similar feature is present in elemental vanadium, 
where it has been attributed to a van Hove singularity at the 
boundary of the V metal Brillouin zone \cite{Scherz}. 
The inset shows clearly that the oxygen contamination affects the V L edge. 
Additional structures show up at the high energy side of the 
V L$_3$ peak, and the V L$_2$ structure is broadened. 
In this sense, at least the surface of the slightly Ti-doped sample 
previously studied \cite{Learmonth} was contaminated by oxygen.  

The V~L$_{2,3}$ maxima for the \textbf{E}$\parallel$\textbf{c} 
polarization are at slightly higher photon energy than for the 
\textbf{E}$\perp$\textbf{c} polarization. 
The shift is about 80~meV in the hPM phase. 
The peak position for V~L$_3$ with \textbf{E}$\perp$\textbf{c} excitation is 516.15~eV.

\begin{figure}
\begin{center}
\includegraphics [width=8.5cm,angle=0]{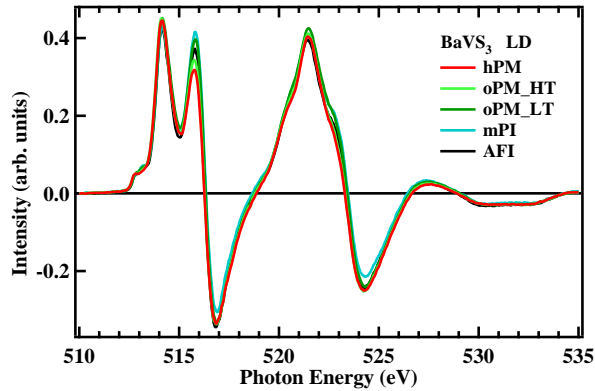}
\caption{\label{fig:BaVS3_LD} (Color online) LD as a function of temperature. 
Red/light-green/dark-green/blue/black curves correspond respectively to the 
hPM/oPM-HT/oPM-LT/mPI/AFI temperature ranges.}
\end{center}
\end{figure}

In order to show more clearly the polarization dependence of the XAS 
intensity, we subtracted $I_{||}$ (\textbf{E}$\parallel$\textbf{c} spectra) 
from $I_{\perp}$ (\textbf{E}$\perp$\textbf{c} spectra).
This linear dichroism (LD) curve is presented in Fig.~\ref{fig:XAS_pol_dep} 
for the hPM phase. 
The change from positive to negative LD simply reflects the shift in the main XAS peak 
according to the polarization of the incident photons. This derivative-like shape hardly 
changes except for differences in amplitude as a function 
of the temperature or phase (Fig.~\ref{fig:BaVS3_LD}). 

When cooling to below $T_S$ = 240~K (hPM to oPM), the shift increases to 110~meV. 
As a simplification we show the average spectrum taken at 300~K, 275~K and 250~K to 
represent the hPM phase, what we term the high temperature orthorhombic phase (oPM-HT) 
is taken as the average of spectra recorded at 225~K, 200~K, 175~K, and 150~K, and the 
low temperature orthorhombic phase (oPM-LT) spectrum is the average of data taken at 125~K, 
100~K and 75~K. The mPI spectrum was recorded at 50~K and the AFI spectrum at 25~K. The shape 
of the main V L$_3$ XAS peak remains practically unchanged but the 
amplitude of the LD signal at 515.8~eV changes reflecting very small variations in the 
relative intensity of the $I_{||}$ and $I_{\perp}$ maxima. 
Another peak in the LD is observed at 514.2~eV, coinciding with the shoulder on the 
XAS data. Its amplitude is positive and remains practically unchanged across the range 
of measurements.

\begin{figure}
\begin{center}
\includegraphics [width=8.5cm,angle=0]{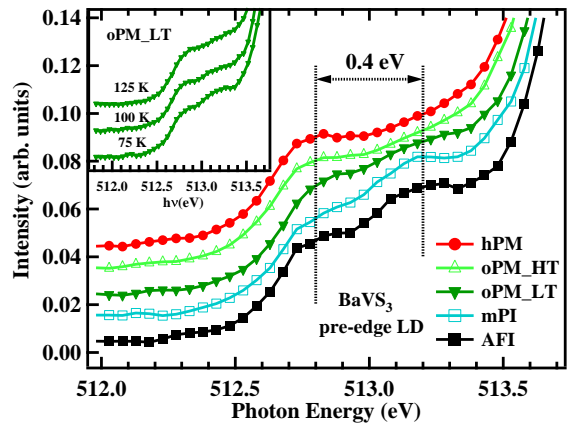}
\caption{\label{fig:BaVS3_LD_PreEdge} 
(Color online) LD changes with temperature in the pre-edge region. 
The spectra are shifted vertically for clarity.
Inset shows details of the temperature changes in the oPM-LT range.}
\end{center}
\end{figure}

An essentially equivalent situation is observed for the L$_2$ edge. 
Here structures are not well resolved. This can be explained by the difference in 
lifetime broadening of the two edges, which is  0.78~eV (0.28~eV) for the V L$_2$ (L$_3$) 
\cite{Campbell2001}.\\
A detailed examination of the low energy part of the spectra, brings to 
light structure at 512.8~eV. The LD data are shown in Fig.~\ref{fig:BaVS3_LD_PreEdge}. 
As the temperature is lowered to the oPM-LT temperature range the intensity is increased 
at 513.2~eV, where a relatively strong extra peak is clearly present  
in both the mPI and AFI phases.

\begin{figure}
\begin{center}
\includegraphics [width=8cm,angle=0]{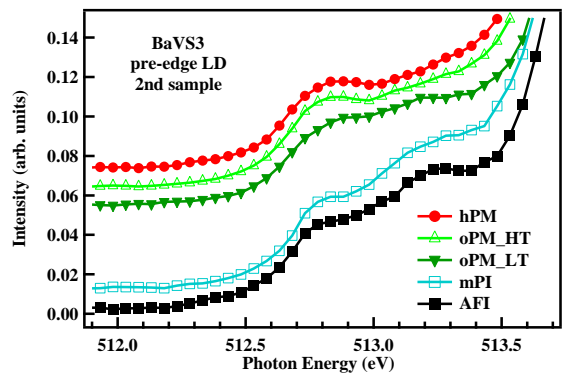}
\caption{\label{fig:BaVS3_LD_sec} 
(Color online) LD changes with temperature in the pre-edge region for the sample 
from another preparation is essentially the same.}
\end{center}
\end{figure}

It should be underlined that we have observed the same low temperature behavior 
for $two$ samples from different preparations. 
Fig.~\ref{fig:BaVS3_LD_sec} shows the LD for the sample from the second batch 
demonstrating that the data are highly reproducible.

\section{Discussion}
\label{sec:Disc}

An elaborate theoretical interpretation of the BaVS$_3$ XAS spectrum is not presently 
available because of its structural complexity. We will therefore discuss our data in terms of 
a simplified analysis of orbital symmetries to help assess what information can be extracted 
from the XAS polarization dependence, along with our present knowledge of the physical 
properties of the material and the insight that can be gained from recent LDA-DMFT 
calculations \cite{Lechermann2007}. 

First we need to calculate the polarization dependence of the $2p$ to $3d$ dipole transition 
in the material to assess the polarization effects.

\begin{figure}[h]
\begin{center}
\includegraphics [width=8.5cm,angle=0]{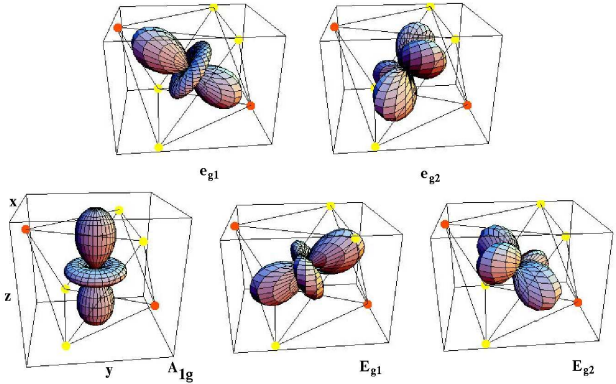}
\caption{\label{fig:Orbitale} (Color online) 
Five BaVS$_3$ V 3d orbitals corresponding to Eq.~\ref{d-orbitals}: 
two e$_g$ (upper panels) and three t$_{2g}$ (lower panels). 
They are inserted in the orthorhombic 
phase deformed octahedra, with two inequivalent sulfur atoms, S$_1$ (orange) 
and S$_2$ (yellow). Axes (x,y,z) correspond to 
(a,b,c) crystallographic axes in the orthorhombic phase.}
\end{center}
\end{figure}

Due to the inclination of the V-S octahedra in BaVS$_3$,
it is convenient to present its V~$3d$ states as a linear 
combination of cubic $d$ orbitals (Eq.~\ref{d-orbitals} 
and Fig.~\ref{fig:Orbitale}):
\begin{eqnarray}
\label{d-orbitals}
\lefteqn{A_{1g}=d_{z^2}}\nonumber\\
   E_{g_1} &=& - 
\sqrt{\dfrac{2}{3}}d_{x^2-y^2}+\sqrt{\dfrac{1}{3}}d_{yz}\nonumber\\
   E_{g_2} &=&  \sqrt{\dfrac{2}{3}}d_{xy}-\sqrt{\dfrac{1}{3}}d_{xz}\\
   e_{g_1} &=& 
\sqrt{\dfrac{1}{3}}d_{x^2-y^2}+\sqrt{\dfrac{2}{3}}d_{yz} \nonumber\\
   e_{g_2} &=&  
\sqrt{\dfrac{1}{3}}d_{xy}+\sqrt{\dfrac{2}{3}}d_{xz},\nonumber
\end{eqnarray}
The 2$p$ $\rightarrow$ 3$d$ transition probabilities, 
according to dipolar selection rules, are given by

\begin{equation}
\label{transitions}
P_{d} = \sum_{i=1}^3 
\left| \langle p_i |\textbf{E} \cdot \textbf{r}| d \rangle \right|^2
\end{equation}
where $p_i$ represents three V 2$p$ orbitals and 
$d$ can be any of the five V $3d$ orbitals: $A_{1g}$, $E_{g1}$, $E_{g2}$, $e_{g1}$, $e_{g2}$. 
$\textbf{E} \cdot \textbf{r}$ is the dipolar 
transition operator. 

The polarization dependent transition probabilities, 
$P^{\perp}$ for \textbf{E}$\perp$\textbf{c} 
and $P^{\parallel}$ for \textbf{E}$\parallel$\textbf{c}, 
are the following:\\
- Transitions to the higher energy $e_g$ states are not polarization 
dependent. For both polarizations their probability is 
$P_{e_g}^{\perp}$ = $P_{e_g}^{\parallel}$ = 0.4.\\
- Transitions to the $A_{1g}$ states are strongly polarization 
dependent, $P_{A_{1g}}^{\perp}$ = 0.1 and 
$P_{A_{1g}}^{\parallel}$ = 0.4.\\
- Transitions to $E_{g}$ states are also polarization dependent, $P_{E_{g}}^{\perp}$ = 0.5 and 
$P_{E_{g}}^{\parallel}$ = 0.2.
Thus the total polarization dependent absorption intensity as a function of energy can be 
written as:

\begin{eqnarray}
\label{intensities}
I_{\perp, \parallel}(\epsilon) \propto 
n_{A_{1g}}(\epsilon) P_{A_{1g}}^{\perp, \parallel} 
+ [n_{E_{g1}}(\epsilon) + n_{E_{g2}}(\epsilon)] 
P_{E_{g}}^{\perp, \parallel} \nonumber \\
+ [n_{e_{g1}}(\epsilon) + n_{e_{g2}}(\epsilon)] 
P_{e_{g}}^{\perp, \parallel},
\end{eqnarray}

where $n_d(\epsilon)$ are unoccupied spectral function weights at a particular 
final (core-excited) state energy $\epsilon$.

As the transitions to the $e_g$ states are equiprobable for both 
polarizations, the linear dichroism should be zero for excitations to states with 
this symmetry. The difference in the transition probabilities to $A_{1g}$ is equal 
to that of each individual $E_{g}$ orbital but of opposite sign. 
Thus an increase in the population of $A_{1g}$ states will tend to reduce the LD or 
give a negative signal while an increase in the $E_{g}$ population will tend to 
increase the LD when defined as $(I_{\perp}~-~I_{\parallel})(\epsilon)$. 

\begin{figure}
\begin{center}
\includegraphics [width=9cm,angle=0]{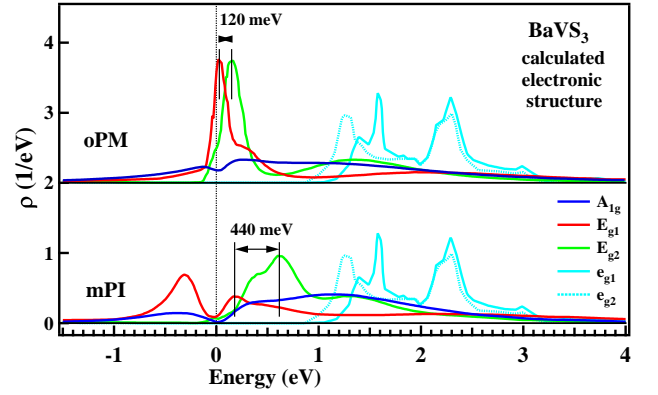}
\caption{\label{fig:Frank_LDA_DMFT}
Local spectral functions from LDA+DMFT calculations of the $t_{2g}$ states in the oPM phase 
where all V sites are equivalent (upper panel) and mean value of local spectral functions for 
four inequivalent V-sites for the $t_{2g}$ states in the mPI phase (lower panel). 
The LDA density of $e_{g}$ states shown in both panels is the one calculated for the oPM phase, 
as the calculation for the mPI phase is not available. Fermi level energy is set to 0. 
LDA and DMFT data have been adapted from the Figs. 2, 7 and 12 of 
Ref.\onlinecite{Lechermann2007}.}
\end{center}
\end{figure}

Summarizing our observations, little difference in the polarization dependence can be 
detected across the T$_S$~=~240~K hPM to oPM phase transition despite the symmetry break 
which lifts the degeneracy of the two $E_{g}$ and two $e_{g}$ states as shown in 
Fig.~\ref{fig:Frank_LDA_DMFT} (upper panel). 
Lowering the temperature below 150~K (oPM-LT) induces an increase in the LD amplitude at 
515.8~eV, while the intensity starts to build up at 513.2~eV, 
where a distinct peak appears below T$_{MI}$ = 69~K. 
When the temperature is further reduced below the 
mPI to AFI ($T_\chi$ = 30~K) phase transition the intensity at 515.8~eV again drops 
and the 513.2~eV peak intensity strengthens. 
This could mean that spin ordering has the effect of transfering one part of 
unoccipied $E_{g}$ states to lower energy.

The increase in LD amplitude at 513.2 eV and 515.8~eV 
across the oPM-HT to oPM-LT temperature range 
might be interpreted as an increase of the $E_g$ weight in the corresponding small energy range. 
As the magnetic properties are related to the $E_g$ states, this interpretation is in line 
with the energy dependence of the ground state magnetic reflection at the (0.226 0.226 0.033) 
wave vector which shows two V L$_3$ peaks at similar positions \cite{Leininger}. 
On the other hand, in the oPM-LT temperature range, there is a formation of CDWs in chains,
as shown by pretransitional diffuse lines in x-ray diagrams \cite{Fagot1}. 
It is a dynamical effect, with a typical lifetime of 
10$^{-12}$~s \cite{PougetKMO}, which can induce local structure modifications via 
electron-phonon interaction. 
As the time scale of XAS is determined by the V L$_3$ edge core-hole-lifetime 
of 10$^{-15}$~s, it can probe short-time modifications of unoccupied states related to CDW fluctuations. 
Similar dynamical local structural change is observed at the Ni K-edge NEXAFS of PrNiO$_3$ \cite{Acosta} system.
 
Over the oPM-HT temperature range all V sites are equivalent by symmetry, while in the mPI phase 
there are four inequivalent V-sites in the monoclinic unit cell \cite{FagotS}. 
Resonant diffraction measurements at the V K edge pointed to the 
possibility of an orbital ordering of $A0E0$ type ($A_{1g}$, equal, 
$E_{g1}$, equal) accompanying the CDW formation \cite{Fagot2}. 
Collective intersite orbital excitations measured in the dielectric signal 
support this hypothesis \cite{Ivek}. 
DMFT calculations confirm that the four inequivalent V-sites have different spectral 
functions but propose rather an $EE00$ occupation pattern \cite{Lechermann2007}. 
In terms of $unoccupied$ states, two sites have their $E_{g2}$ local spectral function 
maximum closer to E$_F$ and the other pair have the maximum pushed to higher energy, 
giving rise to a broad double peak when the mean value for four inequivalent V-sites 
is presented as in the Fig.~\ref{fig:Frank_LDA_DMFT} (lower panel).

From these considerations we suppose that, below 150~K, the increased LD intensity 
at 513.2~eV and 515.8~eV is related to the strong high energy transfer of a part 
of unoccupied $E_{g}$ states. This shift is itself related to the formation of 
four inequivalent V-sites which accompanyies CDW formation in the pretransitional 
fluctuation regime. 
Possible core-hole effects are difficult to estimate without proper theoretical modeling. 
It is however reasonable to suppose that difference in core hole interaction across the 
phases is negligibly small.

\section{Conclusion}
\label{sec:Concl}
BaVS$_3$ single crystal V L edge x-ray linear dichroism (LD) spectra show 
changes with temperature. 
Below 150~K there is a clear LD intensity increase at 513.2~eV and 515.8~eV. 
Simple symmetry analysis suggests the effect is related to rearrangements in $E_{g}$ 
and $A_{1g}$ states. 
Two V L$_3$ peaks at similar energies are observed in the energy scan  
of the ground state (0.226 0.226 0.033) reflection measured by the resonant magnetic 
x-ray scattering \cite{Leininger}. 
This confirms that the observed increase of the LD intensity is related to magnetically 
active $E_g$ states.
DMFT calculations \cite{Lechermann2007} predict rearrangements of $E_{g}$ states in the mPI phase, 
together with CDW-related formation of four inequivalent vanadium sites with 
different $E_{g}$ and $A_{1g}$ spectral functions. 
Our data show that the $E_{g}$ rearrangements start at higher temperatures 
with the formation of CDWs in chains, affecting $A_{1g}$ states. 
We expect these experimental results to help ongoing efforts to model this complex material as 
they highlight the need to consider $A_{1g}$ and $E_{g}$ states as a whole to 
understand the physical properties of this unusual compound.

\acknowledgments
Fruitful discussions with P. Foury, J.-P. Pouget, S. Tomi\'{c}, 
F. Lechermann, S. Bari\v{s}i\'{c}, I. Kup\v{c}i\'{c}, M. Grioni, P. Ben\v{c}ok, 
M.-A. Arrio, P. Sainctavit and Y. Joly are gratefully acknowledged. 
The work in Lausanne was supported by the Swiss National Science Foundation 
and its NCCR ``MaNEP''.


\begin{thebibliography}{99}
\bibitem{Gardner} Gardner R, Vlasse M and Wold A 1969 {\it Acta Crystallogr.} B \textbf{25} 781

\bibitem{Mihaly} Mih\'aly G, K\'ezsm\'arki I, Z\'amborszky F, Miljak M, 
Penc K, Fazekas P, Berger H and Forr\'o L 2000 {\it Phys. Rev.} B \textbf{61} R7831 

\bibitem{Fagot1} Fagot S, Foury-Leylekian P, Ravy S, Pouget J-P and Berger H 2003 
{\it Phys. Rev. Lett.} \textbf{90} 196401

\bibitem{Inami} Inami T, Ohwada K, Kimura H, Watanabe M, Noda Y, Nakamura H, Yamasaki T, 
Shiga M, Ikeda N and Murakami Y 2002 {\it Phys. Rev.} B \textbf{66} 073108

\bibitem{H_NakamuraJPSJ} Nakamura H, Yamasaki T, Giri S, Imai H, Shiga M, Kojima K, 
Nishi M, Kakurai K and Metoki N 2000 {\it J. Phys. Soc. J.} \textbf{69}  2763

\bibitem{Leininger} Leininger Ph, Ilakovac V, Joly Y, Schierle E, Weschke E, Bunau O, Berger H, 
Pouget J-P and Foury-Leylekian P 2011 {\it Phys. Rev. Lett.} \textbf{106} 167203

\bibitem{Mattheiss} Mattheiss L 1995 {\it Solid State Commun.} \textbf{93} 791

\bibitem{Whangbo} Whangbo M-H, Koo H-J, Dai D and Villesuzanne A 2002 
{\it J. Solid State Chem.} \textbf{165} 345

\bibitem{Lechermann} Lechermann F, Biermann S and Georges A 2005 {\it Phys. Rev. Lett.} \textbf{94} 166402

\bibitem{NakamuraPRL} Nakamura H, Imai H and Shiga M 1997 {\it Phys. Rev. Lett.} \textbf{79} 3779

\bibitem{Berger} Kuriyaki H, Berger H, Nishioka S, Kawakami H, Hirakawa K and L\'evy F A 1995 
{\it Synth. Met.} \textbf{71} 2049

\bibitem{Ilakovac} Ilakovac V, Kralj M, Pervan P, Richter M C, Goldoni A, Larciprete R, 
Petaccia L and Hricovini K 2005 {\it Phys. Rev.} B \textbf{71} 085413

\bibitem{Abbate} Abbate M, de Groot F M F, Fuggle J C, Ma Y J, Chen C T, Sette F, 
Fujimori A, Ueda Y and Kosuge K 1991 {\it Phys. Rev.} B \textbf{43} 7263

\bibitem{Park2000} Park J-H , Tjeng L H, Tanaka A, Allen J W, Chen C T, Metcalf P, Honig J M, 
de Groot F M F and Sawatzky G A 2000 {\it Phys. Rev.} B \textbf{61} 11506

\bibitem{Scherz} Scherz A, Wende H, Baberschke K, Min\'ar J, Benea D and Ebert H 2002 
{\it Phys. Rev.} B \textbf{66} 184401

\bibitem{Learmonth} Learmonth T, Glans P A, Guo J H, Greenblatt M and Smith K E 2010 
{\it J. Phys.: Condens. Matter} \textbf{22} 025504

\bibitem{Campbell2001} Campbell J L and Papp T 2001 
{\it At. Data and Nucl. Data Tables} \textbf{77} 1-56

\bibitem{Lechermann2007} Lechermann F, Biermann S, and Georges A 2007 
{\it Phys. Rev.} B \textbf{76} 085101

\bibitem{PougetKMO} Pouget J-P, Hennion B, Escribe-Filippini C and Sato M 1991 
{\it Phys. Rev.} B \textbf{43} 8421

\bibitem{Acosta} Acosta-Alejandro M, Mustre de Le\'on J, Medarde M, Lacorre Ph, 
Konder K, and Montano P A, 2008 {\it Phys. Rev.} B \textbf{77} 085107

\bibitem{FagotS} Fagot S, Foury-Leylekian P, Ravy S, Pouget J P, Anne M, Popov G, Lobanov M V, 
Greenblatt M 2005 {\it Sol. St. Sci.} \textbf{7} 718

\bibitem{Fagot2} Fagot S, Foury-Leylekian P, Ravy S, Pouget J-P, Lorenzo \'E., Joly Y, 
Greenblatt M, Lobanov M V, and Popov G 2006 {\it Phys. Rev.} B \textbf{73} 033102

\bibitem{Ivek} Ivek T, Vuleti\'c T, Tomi\'c S, Akrap A, Berger H and 
Forr\'o L 2008 {\it Phys. Rev.} B \textbf{78} 035110

\end{thebibliography}
\end{document}